# A BeppoSAX and ROSAT view of the RCW86 supernova remnant

**F. Bocchino[1,2], J. Vink[3], F. Favata[1], A. Maggio[2], and S. Sciortino[2]**

[1] Astrophysics Division, Space Science Department of ESA, ESTEC, Postbus 299, 2200 AG Noordwijk, The Netherlands
[2] Osservatorio Astronomico di Palermo, Piazza del Parlamento 1, 90134 Palermo, Italy
[3] Astrophysikalisches Institut Potsdam, An der Sternwarte 16, D-14482 Potsdam, Germany

**Abstract.** We present a spectral analysis of the Northern and Southwestern rim of the RCW86 X-ray shell, pointing out the remarkable differences between these two parts of the same object. In the North, a single temperature Non Equilibrium of Ionization emission model describes the data well, and the derived abundances of O, Ne, Mg, Si, S, Ar and Fe are in agreement with expected metal depletion behind a fast shock moving in the interstellar medium. If the initial explosion energy was $\sim 10^{51}$ erg, the derived distance and age are $1.18^{+0.17}_{-0.16}$ kpc and $1630^{+440}_{-360}$ yr, fully consistent with the association to the historical supernova SN185. The X-ray emission from the SW is described by a two-temperature emission model with $kT_h = 5.7$ keV and $kT_l = 1.0$ keV. There is evidence of overabundant metals in the high-temperature component, thus indicating the presence of ejecta. In this case, a Type Ia SN is to be preferred over a more energetic Type II event, which would imply a more distant and older remnant fully in its Sedov phase.

**Key words:** supernovae: general; ISM: individual object: RCW86; ISM: supernova remnants; X-rays: ISM

## 1. Introduction

The supernova remnant RCW86 (also known as G315.4–2.3 and MSH14–63) is a complete shell in radio (Kesteven & Caswell 1987), optical (Smith 1997) and X-rays (Pisarski et al. 1984), with a nearly circular shape and a 40' diameter. It has received substantial attention because of the longstanding issue of its correlation with SN185, the first historical galactic supernova. However, this connection is based on circumstantial evidence (Clark & Stephenson 1977), and a recent reinterpretation of the Chinese records has even raised some doubt whether the events described really refer to a supernova, rather than a comet (Chin & Huang 1994).

*Send offprint requests to:* F.Bocchino (fbocchin@astro.estec.esa.nl)

Related to the issue is the distance to the remnant. For RCW86 to be the remnant of AD 185, it has to be closer than 2 kpc (Pisarski et al. 1984) with a distance around 1 kpc favored (e.g., Long & Blair 1990). However, most distance estimates either assume that RCW86 is SN185, or imply a distance much larger than 2 kpc. This larger distance dates back to Westerlund (1969), who suggested a possible connection between RCW86 and an OB association at 2.5 kpc. This association is consistent with the distance based on the $\Sigma - D$ relation[1] ($\sim 3$ kpc) and a recent determination based on the systematic velocity of the nebula as measured in $H_\alpha$, which implies a distance of 2.8 kpc (Rosado et al. 1996). However, a distance around 3 kpc is not universally accepted, since the $\Sigma - D$ method is unreliable (Green 1991; Strom 1994), and using the systematic velocity to measure the distance relies upon the assumption that RCW86 is the result of a Type II explosion (Rosado et al. 1996).

A number of features make RCW86 a very interesting remnant in its own right. It displays a large contrast in densities (Leibowitz & Danziger 1983; Pisarski et al. 1984; Petruk 1999) and recently Vink et al. (1997), using ASCA data, showed that two very distinct X-ray spectra are present. One type of spectra is soft and is associated mostly with the radiative shocks making up the knee of RCW86 (c.f. Smith 1997). The hard spectra are associated with the rest of the remnant. The spatial separation of the two components implies that the morphology of RCW86 changes dramatically when going from photon energies around 1 keV to energies above ~3 keV. Some of the peculiar features of this remnant were interpreted by Vink et al. (1997) as due to a partial interaction with a cavity wall. In this view, the non-radiative shocks are still moving within the cavity whereas the bright, radiative shocks are due to encounters between the remnant and the cavity wall.

Vink et al. (1997) also pointed out that the analysis of the ASCA hard spectra of RCW86 yielded strong under-abundances of several elements ($< 0.25$ solar) at several locations of the shell. However, it is not clear if these abundances represent the true abundances, since they are not consistent either with models of grain destructions behind the shock, or with the presence of reverse shocks in the ejecta.

---

[1] This empirical relation links the (radio) surface brightness of a SNR ($\Sigma$) with its real diameter ($D$).

In this paper we describe the BeppoSAX observations of RCW86. Like ASCA, BeppoSAX, is also capable of spatially resolved spectroscopy. The imaging instruments onboard BeppoSAX have a spatial resolution better than those on ASCA and, moreover, span a wider energy range. The BeppoSAX observations also cover the Northwestern part of the remnant, which was not observed by ASCA. In addition, we analyze archival ROSAT Poisision Sensitive Proportional Counter (PSPC) data, and provide new estimates of the metal abundances in RCW86. We will also attempt to interpret the data in the light of a shock expanding in a not homogeneous medium, finding evidence of shocked ejecta in the Southwestern (SW) part of the shell. We will use the Sedov model in conjunction with the X-ray emission of the Northern (N) rim to derive a new and independent estimate of the distance towards this remnant.

In Sect. 2 we describe the data. In Sect. 3, we show the results of spatial and spectral analysis of the SW and N rim of RCW86, while in Sect. 4 we discuss the interpretation of the results. Sect. 5 summarizes our findings.

## 2. Observations

### 2.1. BeppoSAX

RCW86 was observed by the BeppoSAX satellite (Boella et al. 1997a) twice: on February 28th, 1997, the SW part of the shell ("the knee") was pointed at RA=$14^h$ $40^m$ $42.8^s$ and DEC=$-62^d$ $40^m$ $45^s$ (J2000); the NW rim was observed on January 12th, 1998, at RA=$14^h$ $41^m$ $59.5^s$ and DEC=$-62^d$ $13^m$ $00^s$ (J2000). The two observations were processed with the SAXDAS software version 1.6.1 applying the standard corrections and event selections. Table 1 shows the properties of the resulting event lists. While RCW86 has a strong signal both in the LECS (Parmar et al. 1997) and MECS (Boella et al. 1997b) from 0.12 to 10.5 keV, we do not detect any signal in the HPGSPC (Manzo et al. 1997) and only a very weak signal in the PDS (Frontera et al. 1997) in the N pointing. All the 3 MECS detectors were available at the time of the SW observation, while the observation of the NW part was made after the MECS1 detector had failed.

### 2.2. ROSAT

There are 6 ROSAT (Trumper 1983) observation of RCW86. times. In the present work, we used the 4 public archive observations listed in Table 2, which have been obtained between 1992 September and 1995 August. Vink et al. (1997) used the same datasets to derive an hardness ratio map of this remnant, but no other reference to spectral analysis of these data have been found.

**Table 1.** BeppoSAX observations of RCW86

| Locat. | LECS | | MECS | |
|---|---|---|---|---|
| | $T_{exp}$ (sec) | Rate (cnt s$^{-1}$) | $T_{exp}$ (sec) | Rate (cnt s$^{-1}$) |
| RCW86 (N) | 9993 | 0.41 | 23641 | 0.64 |
| RCW86 (SW) | 1836 | 1.22 | 9435 | 1.40 |

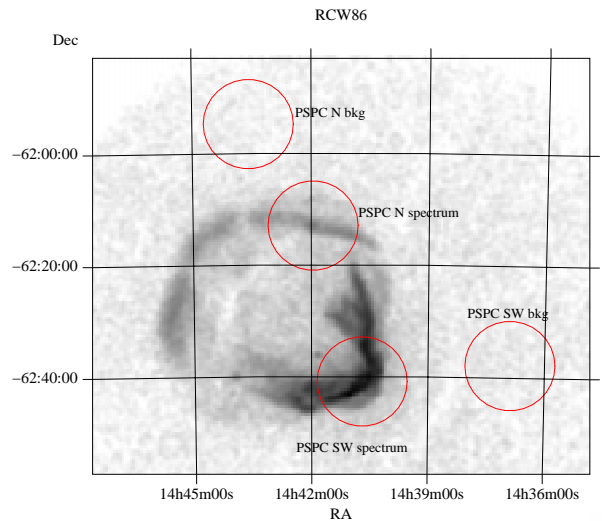

**Fig. 1.** *ROSAT* PSPC (0.1-2.4 keV) image of RCW86. The 8' radius circles represent the extraction region of the source and background spectrum for the PSPC spectral analysis.

Fig. 1 shows the PSPC image of RCW86 with the regions used in the collection of source and background spectra. The background spectra have been corrected for the vignetting effect, using the average exposure time of pixels inside the background and source regions.

## 3. Results

### 3.1. Spatial analysis

In Figs. 2 and 3 we show the *ROSAT* HRI, LECS and MECS images of the SW and N filaments of the RCW86 shell. The LECS image is restricted to the 0.1–2.0 keV band, while the MECS image is restricted to the 2.0–10.5 keV band. The HRI image is in counts per 8 arcsec pixel and smoothed with a gaussian of 1.5 pixel $\sigma$. The BeppoSAX images have been corrected for vignetting effects, have a pixel size of 32 arcsec, and have also been smoothed with a Gaussian of 1.5 pixel $\sigma$. The Point Spread Function (PSF) of the SAX instruments (3.0' at 1.5 keV for the LECS and 2.5' at 6.4 keV for the MECS, 80% of encircled energy radius) does not allow the structure of the X-ray filaments to be traced, which is instead clearly visible in the HRI image. Nevertheless, significant differences of the emission is evident down to the $\sim 1'$ angular scale.

**Table 2.** Public ROSAT observations of RCW86

| Locat. | Seq[a] | T$_{exp}$ (sec) | Coord (J2000) |
|---|---|---|---|
| RCW86 (N) | 500076h | 25692 | $14^h\ 42^m\ 55.0^s\ -62^d\ 12^m\ 36^s$ |
| RCW86 (SW) | 500077h | 8480 | $14^h\ 40^m\ 55.0^s\ -62^d\ 37^m\ 48^s$ |
| RCW86 (N) | 500078p | 9348 | $14^h\ 42^m\ 55.0^s\ -62^d\ 18^m\ 00^s$ |
| RCW86 (SW) | 500079p | 4717 | $14^h\ 41^m\ 26.0^s\ -62^d\ 36^m\ 00^s$ |

[a] "h" indicates HRI observations; "p" indicates PSPC observations.

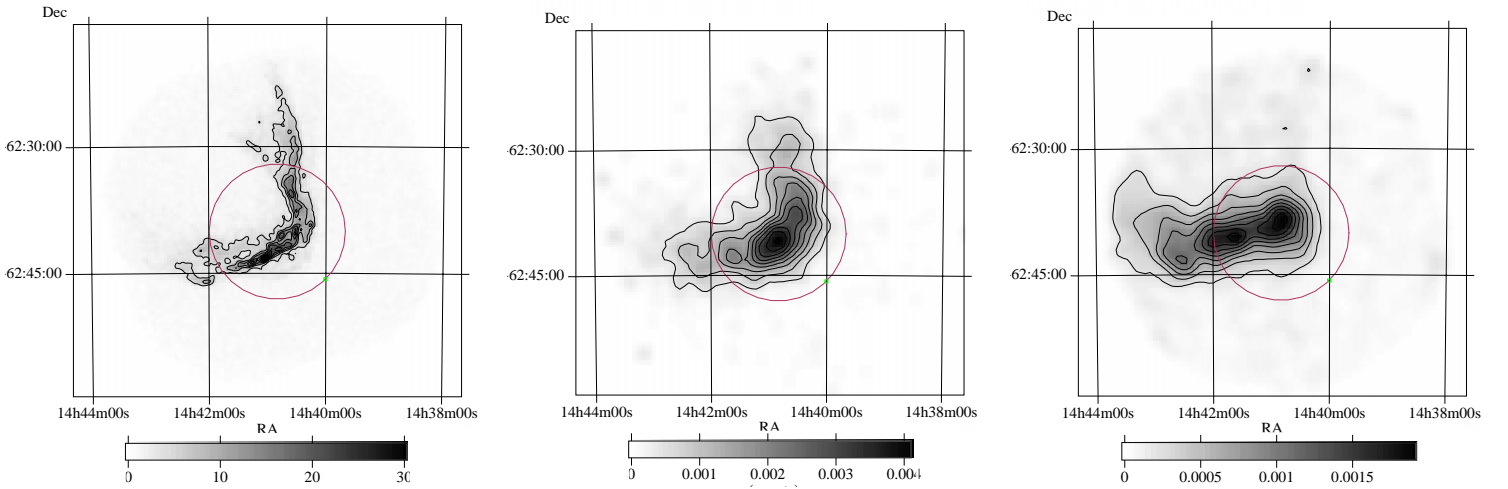

**Fig. 2.** *ROSAT* HRI 0.1-2.4 keV (*left panel*), BeppoSAX LECS 0.1–2.0 keV (*middle panel*) and MECS 2.0–10.5 keV (*right panel*) images of the SW part of the RCW86 shell. Units are counts per $8''$ pixel for the HRI image and counts per second per $32''$ pixel for the BeppoSAX images. Linear contours (10 linearly spaced between 0 and maximum) and grey scale are also shown. The $8'$ radius circles represent the extraction region for the LECS+MECS spectra.

Spectral differences across the filaments are better shown by the hardness ratio maps in Fig. 4. The maps, binned using a pixel size of $\sim 2'$, have been obtained with the LECS 0.1–2.0 keV and MECS 2.0–10.5 keV band data; the value at each pixel has been computed using the formula (HR$= C^{MECS}_{2.0-10.5} - C^{LECS}_{0.1-2.0})/(C^{MECS}_{2.0-10.5} + C^{LECS}_{0.1-2.0})$, where $C$ is the number of counts. The resulting uncertainty in the hardness ratios are between $\pm0.05$ and $\pm0.25$ for $0.5 <$HR$< 1.0$, and between $\pm0.05$ and $\pm0.4$ for HR$< 0.5$. In the SW (Fig. 4, left panel) the soft emission is more external with respect of the hard emission throughout the "knee", except in the East, at RA$=14^h$ $42^m\ 30^s$ and DEC$=-62^d\ 45^m$, where the soft emission is nearly absent and the hard emission is present at the limb of the shell. This position corresponds to the S1b position of Smith (1997), where no [S II] emission is observed, but which does display Balmer-dominated H$\alpha$ filaments. It is interesting to note that while the relation hard emission - Balmer dominated optical filaments seems to hold, the inverse correspondence soft emission - radiative optical filaments is not always fulfilled. In fact, the soft region at the north end of the knee ($14^h\ 41^m$, $-62^d\ 30^m$) corresponds to a region where the [S II] emission is very low and the filaments are mostly Balmer dominated (W1b, W2b and W3b in Smith 1997).

In the N (Fig. 4, right panel) the situation appears to be different, since the soft emission has not a sharp localization and seems more diffuse across the field of view. The soft region around $14^h\ 41^m$ and $-62^d\ 22^m$ is the continuation of the northern and soft end of the knee.

### 3.2. Spectral analysis

#### 3.2.1. The LECS and MECS spectra

LECS and MECS spectra were accumulated in the $8'$ circular regions reported in Figs. 2 and 3. For the LECS, appropriate effective area and response matrix files have been computed with the task LEMAT version 3.5.3, while for the MECS we have used the standard response matrix. In both cases, the effective area files have been computed using the source profile extracted from the HRI image. Background spectra was collected in a BeppoSAX observation of Proxima Centauri, which is $\sim 1.5°$ off RCW86,

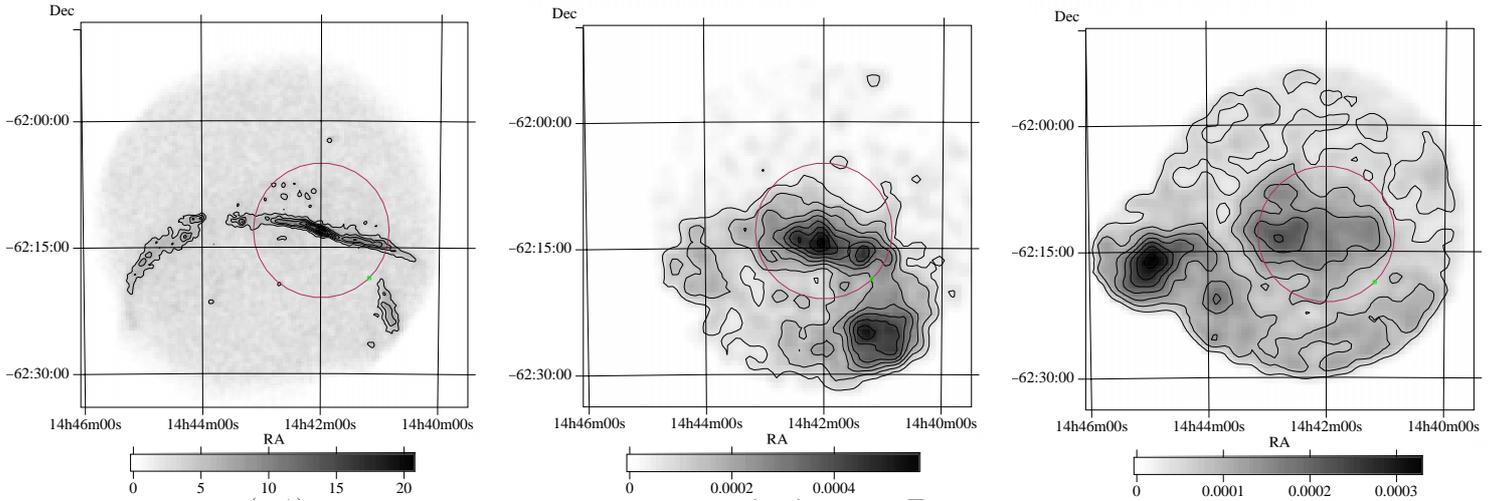

**Fig. 3.** Same as Fig. 2 but for the N part of the RCW86 shell. The asymmetric shape of the field of view, which includes the feature at $14^h$ $45^m$ and $-62^d$ $15^m$, is due to the exclusion of the hot pixels caused by the MECS calibration sources

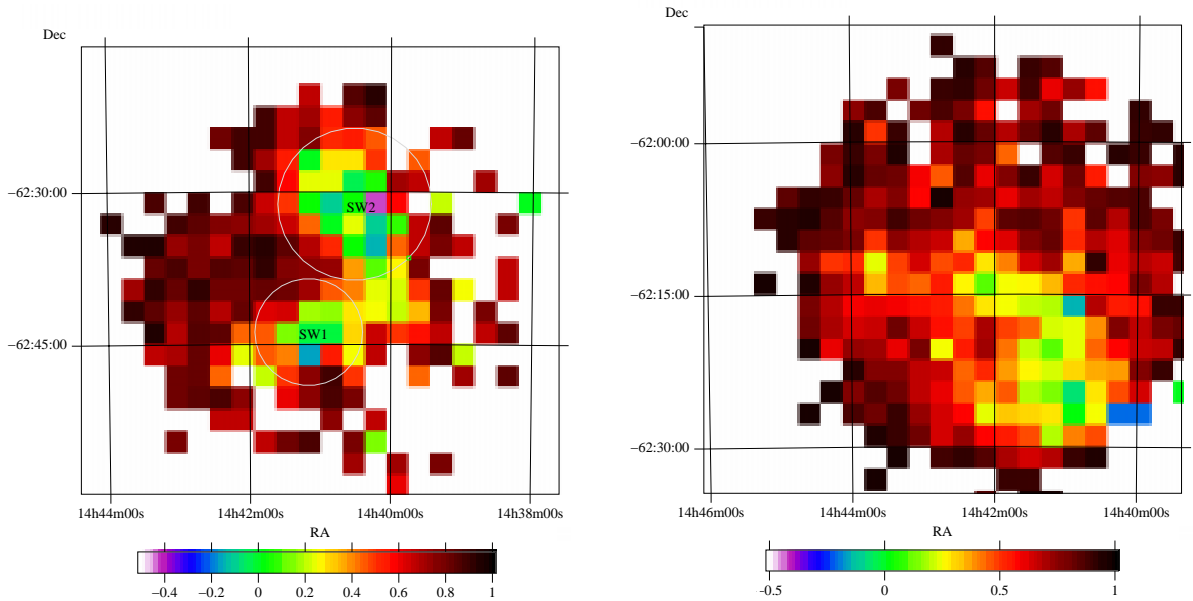

**Fig. 4.** Hardness ratio maps of the South Western part (*left panel*) and of the Northern part of the RCW86 (*right panel*). The maps have been obtained from the LECS 0.1–2.0 keV and MECS 2.0–10.5 keV images (Figs. 2 and 3) binned to a pixel of $2'$; the hardness ratio for each pixel has been computed by the HR= $(C^{MECS}_{2.0-10.5} - C^{LECS}_{0.1-2.0})/(C^{MECS}_{2.0-10.5} + C^{LECS}_{0.1-2.0})$ formula

because there are no suitable background regions in the RCW86 observations. For the LECS we have followed the "background semi-annuli" method of Parmar et al. (1999b) to collect and normalize the background spectrum in the observation of Proxima Centauri, and then we have used it in the RCW86 observation, while for the MECS we have used a similar method, collecting the background in a large annulus around Proxima Centauri, normalizing it to the same area used for the extraction of the source spectrum, and then using it in the RCW86 observation. The "standard background" method also quoted by Parmar et al. (1999b) has proven to be unusable for

RCW86, because at the position of the remnant we expect a non-negligible contribution of the Galactic Ridge X-ray emission and this method would underestimate the background, especially at high energies. The "Scaled ROSAT PSPC all-sky survey" method of Parmar et al. (1999b) has also been discarded since it would provide a background spectrum with too few counts and contaminated by the remnant itself.

The spectra have been rebinned to 1/3 of the effective spectral resolution (FWHM), and then rebinned again to ensure that a minimum of 20 counts are in each channel. The LECS and MECS spectra have been jointly fitted with one temperature and two temperature Non-Equilibrium of Ionization (NEI) emission models included in the SPEX package (Kaastra et al. 1996) plus the absorption component of Morrison & McCammon (1983). The reference abundances are those of Anders & Grevesse (1989), which will be referred in the following as "cosmic". Before fitting, we have multiplied the LECS spectra by the constant 0.87, to take into account the relative normalization between LECS and MECS (see e.g., Favata & Schmitt 1999); we have also verified that varying the multiplication factor yields higher $\chi^2$ values.

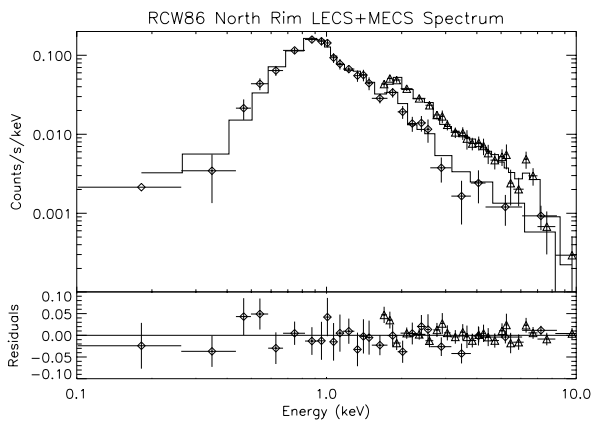

**Fig. 5.** LECS (diamonds) and MECS (triangles) spectrum at the N position of RCW86 with the 1T variable abundances best fit model (continuous line) and percentual residuals

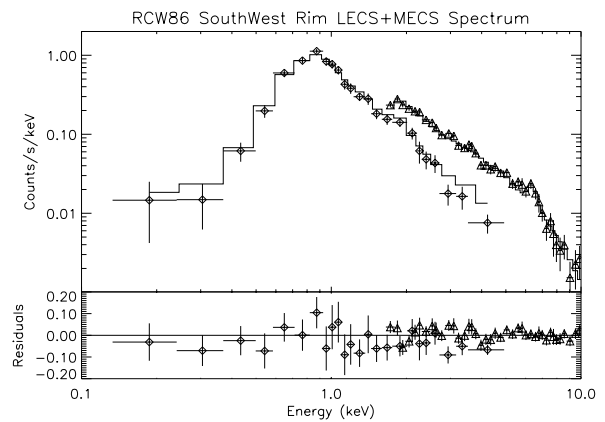

**Fig. 7.** LECS (diamonds) and MECS (triangles) spectrum at the SW position of RCW86 with the 2T variable Z best fit model (cont. line)

One of the dangers of fitting NEI models is to get easily trapped in local minima because of the correlation between the temperature and the ionization time ($\tau$). Including metal abundances as free parameters makes this problem worse. For this reason, we have followed a rigid scheme in fitting our data, starting with a single-component NEI model having fixed abundances, calculating the $\chi^2$ and the corresponding probability, and setting more free parameters only if the fit does not return an acceptable probability. When fitting with variable individual abundances we have chosen to vary only CNO, Ne, Mg, Si, S, Ar and Fe, because these metals mostly affect the BeppoSAX spectra. In particular, the measurement of abundances of some elements is driven by the signature due to line emission in some narrow bands, namely around 1 keV and 6.5 keV for Iron, in the 1.2–1.5 keV band for Mg, 1.8–2.0 keV band for Si, 2.4–2.6 band for S, and the 2.8–3.0 for Ar. Notwithstanding the unaivodable correlation between some of the remaining spectral parameters, the metal abundances of hot thin plasma can be reasonably derived with LECS-MECS spectra, if spectra with a sufficiently high signal-to-noise ratio are available (Favata et al. 1997a; Favata et al. 1998). The results of the one-temperature NEI model fittings obtained in this way are shown in the Table 3.

Neither a 1T model with abundances fixed to cosmic nor a model with variable metallicity (Z) can describe the data at the 10% probability level. This is also true for the variable metal abundances 1T model and the SW spectrum, while the N spectrum is satisfactory reproduced by this model. In Fig. 5, we show the spectrum at the N position, along with the best-fit model of Table 3 and residuals. Since a proper modeling of the X-ray emission from SW cannot be found in the framework of 1T model, we also fitted the SW spectrum with a two-temperature (2T) NEI emission model, obtaining a bad $\chi^2$ using abundances fixed to cosmic, but an acceptable $\chi^2$ using variable metallicity for the two components ($Z_h$ and $Z_l$). The latter result is reported in Table 4, and indicate very different metal abundances for the two components. In Figure 6, we show that $\chi^2$ contours associated to the confidence levels of 68%, 90% and 99% of the two interesting parameters $\log \tau$ and $Z$. We have chosen to represent these two parameters to explore the possible interplay between them, since high abundances may compensate for low $\tau$ values. While the $kT_h = 5.7$ keV component has $Z$ significantly above 1.0, the cooler $kT_l = 1.0$ keV component has $Z = 0.24$, compatible with the best-fit iron abundances of the $kT = 3.6$ keV best-fit model of the N spectrum. Fig. 7 shows the spectrum at the SW position with the best-fit model. Since the measurement of the high-T component abundances is important to the detection of reverse shocks in RCW86, we have verified that our fit has found global minimum rather than a local minimum in the $\chi^2$ space. The best-fit value of the ionization time ($\log \tau = 8.28$ in s cm$^{-3}$) is rather unusual for a SNR, even though it is consistent with the ASCA results obtained by Vink et al. (1997), and indicate a recent interaction between the shock and its environment.

### 3.2.2. ROSAT PSPC spectra

The PSPC spectra have been collected in regions identical in size and position to the regions used for the LECS

**Table 3.** LECS+MECS fitting to single temperature models. The uncertainties are computed using $\Delta\chi^2 = 2.7$, corresponding to 90% confidence level (Lampton et al. 1976)

| Parameter | South Western rim ("the knee") | | | Northern Rim | | |
|---|---|---|---|---|---|---|
| | Fix. Ab. | Var. Z | Var. Ab. | Fix. Ab. | Var. Z | Var. Ab. |
| $\log F^a$ (cm$^{-5}$) | $13.58^{+0.06}_{-0.05}$ | $13.60^{+0.03}_{-0.04}$ | $13.57^{+0.06}_{-0.06}$ | $12.89^{+0.31}_{-0.17}$ | $12.86^{+0.12}_{-0.10}$ | $12.72^{+0.11}_{-0.10}$ |
| kT (keV) | $3.3^{+0.6}_{-0.3}$ | $3.4^{+0.3}_{-0.3}$ | $3.4^{+0.4}_{-0.4}$ | $2.7^{+1.0}_{-0.7}$ | $2.8^{+0.6}_{-0.6}$ | $3.6^{+0.9}_{-0.7}$ |
| $\log \tau$ (s cm$^{-3}$) | $9.75^{+0.05}_{-0.05}$ | $10.18^{+0.08}_{-0.10}$ | $10.18^{+0.12}_{-0.14}$ | $9.77^{+0.19}_{-0.13}$ | $10.15^{+0.10}_{-0.15}$ | $10.28^{+0.12}_{-0.17}$ |
| CNO | 1.0 | =Fe | $0.22^{+0.29}_{-0.11}$ | 1.0 | =Fe | $0.17^{+0.39}_{-0.10}$ |
| Ne | 1.0 | =Fe | < 0.34 | 1.0 | =Fe | < 0.34 |
| Mg | 1.0 | =Fe | $0.38^{+0.16}_{-0.15}$ | 1.0 | =Fe | $0.64^{+0.33}_{-0.25}$ |
| Si | 1.0 | =Fe | $0.33^{+0.10}_{-0.08}$ | 1.0 | =Fe | $0.56^{+0.23}_{-0.17}$ |
| S | 1.0 | =Fe | $0.49^{+0.34}_{-0.21}$ | 1.0 | =Fe | $0.81^{+0.52}_{-0.33}$ |
| Ar | 1.0 | =Fe | < 1.4 | 1.0 | =Fe | < 1.9 |
| Fe | 1.0 | $0.34^{+0.16}_{-0.06}$ | $0.36^{+0.16}_{-0.10}$ | 1.0 | $0.47^{+0.20}_{-0.18}$ | $0.31^{+0.23}_{-0.13}$ |
| $N_H$ (10$^{21}$ cm$^{-2}$) | $4.3^{+0.2}_{-0.2}$ | $2.2^{+0.5}_{-0.5}$ | $1.6^{+1.3}_{-0.7}$ | $5.2^{+0.4}_{-0.4}$ | $3.8^{+0.7}_{-0.8}$ | $1.1^{+1.7}_{-0.5}$ |
| $\chi^2$/dof (%) | 132/61 (<1%) | 91/60 (<1%) | 83/54 (<1%) | 72/48 (1%) | 61/47 (8%) | 38/41 (60%) |

$^a$  $F = \frac{n^2 V}{4\pi D^2}$ is the normalization parameter of the emission model.

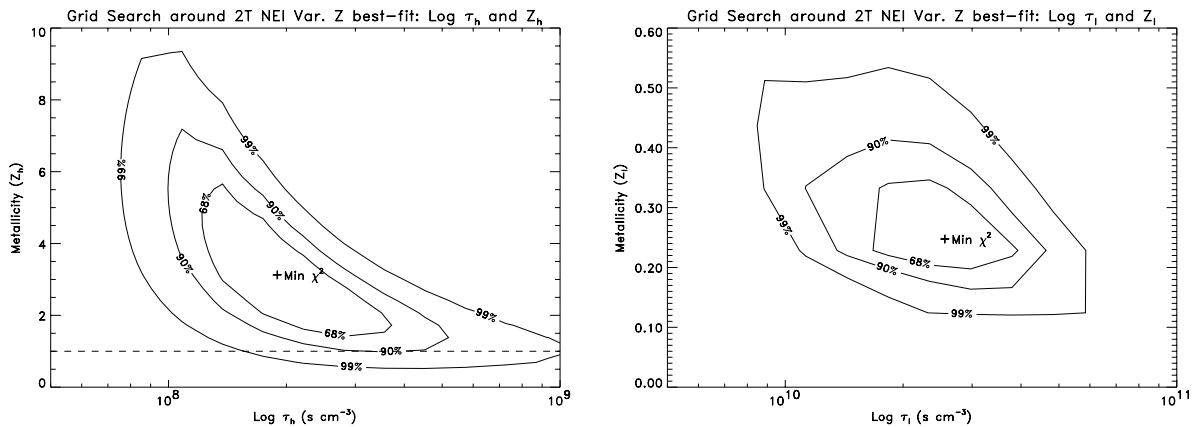

**Fig. 6.** $\chi^2$ contour plots in the $\log \tau - Z$ plane for the high-T component (*left*) and the low-T component (*right*) of the 2T NEI fit with variable metallicities. The 68%, 90% and 99% confidence level contours for two interesting parameters are displayed. The dashed line corresponds to $Z_h = 1$

**Table 4.** LECS+MECS fitting to two temperature models in the South Western region. The uncertainties are computed using $\Delta\chi^2 = 2.7$, corresponding to the 90% confidence level (Lampton et al. 1976)

| Parameter | South Western rim ("the knee") | |
|---|---|---|
| | high-T | low-T |
| $\log F$ (cm$^{-5}$) | $13.35^{+0.16}_{-0.40}$ | $13.89^{+0.24}_{-0.21}$ |
| kT (keV) | $5.7^{+4.0}_{-1.7}$ | $1.0^{+0.4}_{-0.3}$ |
| $\log \tau$ (s cm$^{-3}$) | $8.28^{+0.34}_{-0.23}$ | $10.41^{+0.30}_{-0.27}$ |
| $Z$ | $3.0^{+8.8}_{-1.7}$ | $0.24^{+0.18}_{-0.09}$ |
| $N_H$ (10$^{21}$ cm$^{-2}$) | $2.7^{+1.1}_{-0.9}$ | |
| $\chi^2$/dof (%) | 62/56 (28%) | |

and MECS spectra. The background was collected in the same PSPC image, at an off-axis position free from diffuse emission. The spectra have been rebinned to ensure at least 20 counts in each bin. The spectra have been first fitted with a 1T NEI model with metal abundances both fixed to the cosmic values and jointly varying (free Z), resulting in both cases in a too high $\chi^2$ value. The fact that 1T NEI spectra fail to reproduce the X-ray SNR emission in the PSPC band was also observed for the Vela SNR and there explained with the multi-component nature of the plasma in the interaction regions (Bocchino et al. 1997). In our case, the BeppoSAX fitting results provide us with useful hints about the components present in the plasma, and therefore we have checked whether the model best-fitting the BeppoSAX spectra is able to describe also the PSPC data. Even if the normalization of the thermal components detected with SAX are left free to vary, these fits yielded again an high $\chi^2$, suggesting the presence of an additional PSPC soft component not detected in the BeppoSAX spectra. Therefore, we added a thermal Collisional Ionization Equilibrium (CIE) component to the BeppoSAX-derived components, because a second ther-mal component was also reported by Vink et al. (1997), and we fitted the PSPC spectra. The results, summarized with the other PSPC fitting results in Table 5, shows that the $\chi^2$ improve noticeably with a very soft component at $kT = 0.29$ for the SW spectrum and $kT = 0.17$ keV for the N spectrum. If we fit the LECS+MECS spectra with this supersoft component and the previously derived components with varying normalizations, the derived normalization factors of the soft component are consistent with 0, indicating that only an upper-limit to the PSPC soft component can be obtained on the basis of the BeppoSAX spectra alone.

As Parmar et al. (1999a) summarize, it is not uncommon that the PSPC detects a soft thermal component with $kT \lesssim 0.1$ keV which is not detected in the LECS instrument. This was the case not only for X-ray cosmic background spectrum, but also for the spectrum of some active stars which have been observed with the PSPC and other X-ray and UV detectors (see e.g., Griffiths & Jordan 1998; Brickhouse & Dupree 1998). Parmar et al. (1999a) argue that this might point towards a systematic effect in the low-energy calibration of the PSPC. However, we note that the presence of a supersoft component is in agreement with the idea that RCW86 is expanding in an inhomogenuous medium, as suggested by radiative and nonradiative optical filaments coexisting next to each other (Leibowitz & Danziger 1983; Smith 1997). In this framework, the supersoft component may be the tracer of the interaction of gas with density in between the low density associated with the main shock in the ISM and ejecta, and the high densities associated with the bright optical filaments, as, for instance, is the case for the Vela SNR (Bocchino et al. 2000). Moreover, we also note that the limited spatial resolution of the LECS at low energies (6.1' at 0.28 keV, 80% encircled power radius) is significantly lower than the resolution of the PSPC (33" at 0.3 keV, 80% encircled power radius). This implies that soft energy photons of the supersoft component are spread over a wide area in the LECS image, and even out of our 8' LECS extraction region, while this effect is much smaller in the PSPC. This eventually implies a difficulty to detect super-soft components greater with the LECS than with the PSPC.

## 4. Discussion

Table 6 summarizes the published results obtained by X-ray spectral analysis of RCW86, of which the ASCA observations are the first attempt of spatially resolved analysis. We note that only EXOSAT and GINGA results provide a statistical acceptable description of the RCW86 X-ray emission at the $\chi^2 > \chi_0^2$ probability threshold of 10%. Considering all the other results, the derived values of temperature and abundances, which are two key parameters for the understanding of the SNR and its environment, are often not consistent with each other, and this suggests that this object is still poorly modeled. As a matter of fact, there is still not much agreement on the ultimate nature of the X-ray emission of RCW86, and this is partially due to the fact that this object shows very different properties in the SW and in the N and could not be properly described by fitting its spatially integrated spectrum. This is why in this section we will focus independently on the N and on the SW regions.

### 4.1. Interpretation of the X-ray emission in the N region

We have seen that the X-ray emission of the N filament is well described by a single temperature thermal emission model, with abundances lower than the cosmic values. This result suggests that we have detected the interaction between the blast wave and the ISM. Long & Blair (1990) have imaged and spectroscopically studied the red part of the spectrum of the group of filaments which lie near the center of our extraction region. In particular, they noticed that radiative and non-radiative characteristics are present in the same filament. Smith (1997) suggests that this could be due to a recent encounter with a dense clump. The very soft component we have detected in the PSPC spectra mentioned in Sect. 3.2.2 might be the footprint of this early interaction stage with a ISM clump, as we have discussed therein.

The best-fit temperature ($kT = 3.6$ keV) found in the N filament could be taken as our best guess of the plasma temperature behind the main shock of RCW86. In this case, we can derive the main shock velocity ($v_s = 1720^{+200}_{-180}$ km s$^{-1}$) and compare it with independent estimates, such the one obtained by Long & Blair (1990). In fact, they measured the broadening of the two components in the H$\alpha$ line profile of filaments at N and SW of the remnant and they found $v_s = 800 \pm 130$ km s$^{-1}$, which seems inconsistent with our estimate obtained at the N position. Such a shock would have $kT \sim 1.0$ keV, and would be easily detected by BeppoSAX. However, Long & Blair (1990) noted that the shock velocity derived by the ratio of the broad and narrow component is 1600–1900 km s$^{-1}$, which is consistent with our estimate, but they consider more reliable the value obtained using the width of the broad component, because the calibration of the ratio between the two components sensitively depends on assumptions about the post-shock electron temperature. The two estimates could be reconciled if the shock is encountering a density gradient, e.g. a cavity wall as suggested by Vink et al. (1997), and therefore decelerating. In this case, the X-ray spectra may still be dominated by the low density wind-driven bubble ISM, while the optical filament may arise from the shock in the more dense cavity wall. A similar discrepancy between optical and X-ray derived shock velocities is also reported in the North East of the Cygnus Loop, where the broad H$\alpha$ component gives $v_s = 170$ km s$^{-1}$ (Blair et al. 1999), while ASCA X-ray spectroscopy gives $v_s = 450$ km s$^{-1}$ (Miyata et al. 1994).

**Table 5.** PSPC fitting results on both the SW and N regions. Values with uncertainties (computed with the criterium $\chi^2 < \chi^2_{min} + 2.7$, Lampton et al. 1976) correspond to free parameters

| Parameter | South Western rim ("the knee") | | Northern rim | |
| --- | --- | --- | --- | --- |
| | 1T NEI | 1T CIE + SAX[a] | 1T NEI | 1T CIE + SAX[b] |
| $\log F$ (cm$^{-5}$) | $12.90^{+0.13}_{-0.05}$ | $14.53^{+0.18}_{-0.28}$ | $13.05^{+0.49}_{-0.50}$ | $13.71^{+0.59}_{-0.57}$ |
| T (keV) | $25^{+79}_{-18}$ | $0.29^{+0.03}_{-0.04}$ | $> 0.8$ | $0.17^{+0.01}_{-0.02}$ |
| $\log \tau$ (s cm$^{-3}$) | $10.15^{+0.23}_{-0.20}$ | - | $9.49^{+0.43}_{-0.5}$ | - |
| $N_H$ ($10^{21}$ cm$^{-2}$) | $1.2^{+0.5}_{-0.2}$ | $1.7^{+0.3}_{-0.2}$ | $4.5^{+1.3}_{-2.0}$ | $3.3^{+1.3}_{-1.2}$ |
| $\chi^2$/dof (%) | 60/16 (< 1%) | 23/15 (9%) | 30/16 (2%) | 23/16 (12%) |

[a] These results have been obtained by fitting a 1T CIE model plus two component with their parameters fixed to the best-fit values listed in the Var. Z column of Table 4.

[b] These results have been obtained by fitting a 1T CIE model plus one thermal component with its parameters fixed to the best-fit values derived by the fitting of the SAX Northern rim spectrum, and listed in the Var. Ab. column of Table 3.

**Table 6.** Previous estimates of RCW 86 physical parameters

| Parameter | HEAO1 | | EXOSAT | EINSTEIN | GINGA | ASCA | | |
| --- | --- | --- | --- | --- | --- | --- | --- | --- |
| | CIE | NEI | | | | SWsoft | SWhard | NEhard |
| References[a] | [1] | [1] | [2] | [3] | [4] | [5] | [5] | [5] |
| $kT_h$ (keV) | $5.10^{+0.14}_{-0.14}$ | 5.1 | $3.4^{+0.2}_{-0.2}$ | $1.2^{+0.3}_{-0.1}$ | $4.31^{+0.23}_{-0.23}$ | $0.79^{+0.06}_{-0.06}$ | $1.2^{+0.2}_{-0.2}$ | $2.0^{+0.1}_{-0.2}$ |
| $kT_l$ (keV) | $0.52^{+0.04}_{-0.04}$ | | $0.51^{+0.03}_{-0.03}$ | | | | $5.0^{+\infty}_{-1.3}$ | |
| $\log \tau_h$ (s cm$^{-3}$) | - | $10.20^{+0.03}_{-0.03}$ | - | - | $10.91^{+0.12}_{-0.16}$ | $9.99^{+0.04}_{-0.04}$ | $9.75^{+0.04}_{-0.05}$ | $9.23^{+0.24}_{-0.23}$ |
| $\log \tau_l$ (s cm$^{-3}$) | - | | - | | | | $< 9.41$ | |
| O | | | | | | $0.09^{+0.03}_{-0.03}$ | $0.014^{+0.007}_{-0.006}$ | $< 0.002$ |
| Ne | | | | | | $0.25^{+0.03}_{-0.04}$ | $0.078^{+0.008}_{-0.007}$ | $0.04^{+0.02}_{-0.01}$ |
| Mg | | | | | | $0.12^{+0.03}_{-0.02}$ | $0.06^{+0.01}_{-0.02}$ | $0.03^{+0.06}_{-0.03}$ |
| Si | | $0.7^{+0.1}_{-0.1}$ | | | $0.59^{+0.07}_{-0.07}$ | $0.19^{+0.05}_{-0.05}$ | $0.15^{+20.0}_{-0.05}$ | $0.3^{+0.2}_{-0.2}$ |
| S | | $3.4^{+0.7}_{-0.6}$ | | | $0.62^{+0.07}_{-0.07}$ | $< 0.9$ | $< 0.2$ | $< 1.2$ |
| Fe | | $3.9^{+0.9}_{-1.3}$ | $0.4^{+0.3}_{-0.3}$ | | $0.18^{+0.05}_{-0.05}$ | $0.16^{+0.04}_{-0.04}$ | $0.12^{+0.04}_{-0.03}$ | $2.8^{+5.2}_{-2.7}$ |
| | | | | | | | $2.0^{+1.0}_{-1.0}$ | |
| $N_H$ $10^{21}$ cm$^{-3}$ | $1.1^{+0.3}_{-0.3}$ | $4.4^{+0.3}_{-0.3}$ | $1.3^{+0.2}_{-0.2}$ | $0.3^{+0.2}_{-0.2}$ | $0.0^{+\infty}$ | $2.3^{+0.4}_{-0.3}$ | $1.6^{+0.4}_{-0.3}$ | $1.3^{+0.4}_{-0.3}$ |
| $\chi^2$/dof | 96/60 | 69/53 | 23/19 | 16/8 | 26/30 | $\chi^2_\nu = 4.7$ | $\chi^2_\nu = 1.7$ | $\chi^2_\nu = 1.7$ |

[a] [1] Nugent et al. (1984); [2] Claas et al. (1989); [3] Pisarski et al. (1984); [4] Kaastra et al. (1992); [5] Vink et al. (1997).

The interpretation of the X-ray thermal component in terms of shock propagating in the ISM is also suggested by the observed abundances values. The abundances are in agreement with the value reported in the "NE Hard" region by Vink et al. (1997) (also reported in Table 6), except for O and Mg, for which ASCA reports a value below 0.1; however, the BeppoSAX derived abundances have smaller uncertainties for most of the metals. Vancura et al. (1994) have developed a model of dusty non-radiative shock waves in which they show that metals are locked up in ISM grain and slowly released in the plasma as the grain are destroyed behind the shock. At the shock front, the fraction of metal not depleted onto grains is 0.51 for CNO, 1.0 for Ne, 0.05 for Mg, 0.05 for Si, 0.5 for S, 1.0 for Ar and 0.02 for Fe. Because of grain destruction, the fraction of undepleted material slowly converges toward 1 for all the metals, and this is why we expect to derive metal abundances lower than the cosmic values of

Anders & Grevesse (1989). As a rule of thumb, for the derived abundances to be consistent with a metal depletion model, they should not be lower than the fractions of undepleted metals at the shock front, reported above.

The measured metal abundances in the N filament, reported in Table 3 are all above the expected metal abundances at the shock front (except Ne), and therefore consistent with the dusty shock model. Ne is inconsistent because we expect no depletion in grains, but we observe Ne< 0.34. The measurement of Ne abundances may be however not reliabe with current X-ray instruments, because the Ne lines around 1 keV are embedded in the Fe-L blend. We can calculate the fraction the remaining grain mass ($f_{\rm grain}$) from the derived abundances. $f_{\rm grain}$ is 1 at the shock front and 0 when all the grain have been destroyed and the metal abundances are at their cosmic values. We have

$$f_{grain} = \frac{1 - [x]}{1 - [x]_{sf}} \qquad (1)$$

where $[x]$ is the derived metal abundances and $[x]_{sf}$ is the fraction of metal not depleted onto grains at the shock front, i.e. the expected metal abundance at the shock front. If we consider the abundance of Fe, which is the one with the lowest uncertainty ([Fe]=0.18 − 0.54), we derive by Eq. 1 that the fraction of remaining grain mass is between 0.47 and 0.84. Analogously, the abundance of Si implies a remaining grain mass between 0.21 and 0.64. Combining the two results and following Vancura et al. (1994), we argue that the column swept-up by the shock is $\sim 5 - 10 \times 10^{18}$ cm$^{-2}$, which is equivalent to a distance behind the shock of $\sim 1.5 - 3.3 \times n_{cm^{-3}}^{-1}$ pc, where $n_{cm^{-3}}$ is the post-shock density in cm$^{-3}$.

### 4.2. Sedov analysis of the N emission

Sedov analysis is a powerful tool to derive the remnant characteristic parameters from X-ray spectral analysis. Kassim et al. (1994) reviewed the method, pointining that it could be a valid distance and age estimator, and verified it on a set of SNRs with independent distance estimates. They also stress that the main source of uncertainties in the analysis is the value of the initial explosion energy, which must be assumed and which is poorly known for Type II supernovae. Bocchino et al. (1999) extended the applicability of the method to spatially resolved X-ray spectroscopy of SNRs, and pointed out that the method is valid even in presence of ISM inhomogeneities, provided that the component responsible for the X-ray emission of shocked "inter-cloud" medium is properly identified and used.

The method is based on the mutual dependence of the age ($t_{10^4}$ in units of $10^4$ yr), the pre-shock density ($n_0$ in cm$^{-3}$), the shell radius ($r_{sh}$ in pc), the explosion energy ($E_{51}$ in unit of $10^{51}$ erg) and the shock speed ($v_{s7}$ in units of $10^7$ cm sec$^{-1}$), as given, for instance, by McKee & Hollenbach (1980):

$$t_{10^4} = \frac{14}{v_{s7}^{5/3}} \left(\frac{E_{51}}{n_0}\right)^{1/3}; \quad r_{sh} = \frac{36}{v_{s7}^{2/3}} \left(\frac{E_{51}}{n_0}\right)^{1/3}. \qquad (2)$$

The values of $v_s$ and $n_0$ could be derived by the X-ray results, since

$$v_{s7} = 8.39\sqrt{T_{s7}}; \quad n_0 = \frac{1}{4}n = \frac{1}{4}\sqrt{\frac{F 4\pi}{\theta l}}, \qquad (3)$$

where $T_{s7}$ is in unit of $10^7$ K, $F = \frac{n^2 V}{4\pi D^2}$ is the normalization factor of the spectrum, $\theta$ is the solid angle of the extended source and $l$ is the line of sight extension of the source (Bocchino et al. 1999). The shell radius and the age of the remnant depends weakly on the unknown $l$, and one can safely use a value of few parsec for shell SNR. A very reasonably estimate of $l$ can be derived if we assume that the emission mainly comes in a thin shell behind the shock front, as the Sedov model predicts:

$$l = 2\sqrt{r_{sh}^2 - \left(\frac{11}{12}r_{sh}\right)^2} \sim \frac{4}{5}r_{sh}. \qquad (4)$$

**Table 7.** Results of the detailed Sedov analysis applied to the 1T fitting results of the N region. We have used the apparent shell radius $\Theta = 21'$ reported by Green (1996)

|  | $E_{51}=0.1$ | $E_{51}=1$ | $E_{51}=10$ |
|---|---|---|---|
| $R_{sh}$ (pc) | $2.9^{+0.4}_{-0.4}$ | $7.2^{+1.0}_{-0.9}$ | $18.1^{+2.6}_{-2.4}$ |
| $n_0$ (cm$^{-3}$) | $0.67^{+0.14}_{-0.12}$ | $0.42^{+0.09}_{0.07}$ | $0.27^{+0.05}_{-0.05}$ |
| $D$ (kpc) | $0.47^{+0.07}_{-0.06}$ | $1.18^{+0.17}_{-0.16}$ | $2.97^{+0.43}_{-0.37}$ |
| $t$ (yr) | $650^{+170}_{-150}$ | $1630^{+440}_{-360}$ | $4080^{+1120}_{-900}$ |
| $M_{sw}$ (M$_\odot$) | $1.6^{+1.2}_{-0.7}$ | $15.7^{+12.8}_{-7.1}$ | $157^{+128}_{-71}$ |

This relation assumes that the line of sight is tangential to the inner border of the shell, and that the shell is $r_{sh}/12$ thin. The derived value of $l$ may appear overestimated, but we recall that we are seeing optical and X-ray filaments because of projection of sheet-like structures (Hester 1987), which gives rise to large $l$.

Substituting Eqs. 4 and 3 in Eq. 2, we obtain

$$t_{10^4 yr} = 3 \times 10^4 T_{s7}^{-9/10} E_{51}^{2/5} \left(\frac{F_{cm^{-5}}}{\theta}\right)^{-1/5} \qquad (5)$$

where $T$ and $F$ are given by the X-ray fits and $\theta$ (in steradians) is measured, e.g. in the HRI. In the same way, we can derive the real shell radius, and therefore the SNR distance, because $r_{sh} = 0.29 D_{kpc}\Theta_{arcmin}$ ($\Theta$ is the apparent shell radius, Kassim et al. 1994). Eq. 2 for the shell radius becomes

$$D_{kpc} = 2.3 \times 10^5 \Theta_{arcmin}^{-1} T_{s7}^{-2/5} E_{51}^{2/5} \left(\frac{F_{cm-5}}{\theta}\right)^{-1/5} \qquad (6)$$

In Table 7 we report the quantities computed with Eq. 5 and 6 using the best-fit values and uncertainties of the 1T fit of the Northern Rim listed in Table 3. The swept-up mass ($M_{sw}$) is computed according to Kassim et al. (1993) to check the consistency of the Sedov approach: since we expect that SN ejected masses are in the range 1–5 M$_\odot$ for Type Ia SNe and 5–15 for Type II SNe (Woosley & Weaver 1986), $M_{sw}$ values higher than that indicates Sedov evolution. Given the uncertainties in the $E_{51}$ value, we report the results computed with three different assumptions which covers a wide range of possibilities, namely $E_{51} = 0.1$, 1 and 10. If RCW 86 has been generated by a Type Ia event, then $E_{51}$ should be in the range 0.9-1.5, whereas a larger range is expected in case of Type II events (Kassim et al. 1993). We note that previous estimates of the RCW86 explosion energy point towards $E_{51} = 0.1$ (see Petruk 1999 for a review) but they are derived assuming an age of 1800 yr, and using spatially integrated X-ray data and/or CIE emission models. If $E_{51} = 0.1$ the swept-up mass is in any case lower or comparable with the mass of ejecta, and this would imply that the remnant is still heavily interacting with the ejecta. This is hardly the case, because the derived metal abundances in the N pointings are not above the cosmic abundances. For this reason, we exclude such a low $E_{51}$ value. If we consider the generally accepted value $E_{51} = 1$, $M_{sw}$ is $\gg M_{ej}$ only in case of a Type Ia SN, while both Type Ia and Type II events are allowed by the results obtained by the higher value $E_{51} = 10$.

While we can reasonably exclude $E_{51} = 0.1$, we cannot discriminate between $E_{51} = 1.0$ and $E_{51} = 10$ at this stage, apart from the fact that the canonical value is generally preferred in the literature for the supernovae. If $E_{51} = 1$, the derived age is fully consistent with the RCW86–SN185 association and the distance is in agreement with Ruiz (1981) and Strom (1994), but not with the distance given by Rosado et al. (1996) ($D_{kpc} = 2.8$), which argued against the RCW86–SN185 association, and by Milne (1970) on the basis of the $\Sigma - D$ relation. To reconcile with the proposed association between RCW86 and the OB association proposed by Rosado et al. (1996), we must assume $E_{51} = 10$ and a Type II event.

### 4.3. Reverse shock in the SW region?

The detection of metal abundances above the cosmic values in the hot thermal component of the SW part of RCW86 is in agreement with the ASCA results, and it suggests the possibility that, at this location, the interaction is occurring with the SN ejecta. We note that this is at variance with the results obtained in the N, where there is strong evidence for Sedov-like X-ray emission. This is only apparently surprising, because RCW86 could be expanding in an ISM with large-scale density gradient. In particular, Petruk (1999) has developed a 2-D modeling of the evolution and X-ray emission of RCW86, based on the EXOSAT spatially integrated spectral results reported by Claas et al. (1989), and reported a pre-shock density contrast between the SW and Northeast parts of RCW86 in the range 3.5–4.5. In these conditions, the remnant is expected to have different evolutionary timescales at different shell location, and spatially resolved spectral analysis becomes really necessary to understand the origin of the observed emission.

We can derive the relative filling factor of the two components and their density, using Eq. 4 and 6 of Bocchino et al. (1999), assuming that the two components are roughly in pressure equilibrium, and using the extension of the source along the line of sight given in the previous section ($l \sim 4 r_{sh}/5 = 5.8$ pc). We obtain $f_h = 0.15$, $f_l = 0.85$, $n_h = 3.8$ and $n_l = 3.0$ cm$^{-3}$. We have used a solid angle of the source of $7.6 \times 10^{-6}$ steradians, corresponding to the intersection of our circular extraction region, with a polygon tightly enclosing the shape of the "knee" as seen by the HRI. Moreover, using the distance derived in the previous section, we can derive the X-ray emitting mass of the two components, since $M = m_p n V = m_p n \theta D^2 l$; we find $M_h = 0.8$ M$_\odot$ and $M_l = 3.4$ M$_\odot$ for $E_{51} = 1$ (15 and 68 M$_\odot$, if we use the $E_{51} = 10$ results of Table 7).

If the SN ejecta expands in a uniform medium (as it would be more probable in case of Type Ia SNe) or in stellar wind bubble, we expect that they give rise to an X-ray component which is cooler than the one due to the shock expansion in the ambient medium. According to Chevalier (1982), the temperature of the reverse shock expanding in the ejecta is at most 1.6 times lower than the temperature of the primary blast-wave (corresponding to a power-law density profile $\rho \propto r^{-8}$ for the ejecta), while we observe $T_h/T_l > 2.8$. More recently, Truelove & McKee (1999) presented a detailed semi-analytical model of non-radiative SNR in which they show that the temperature of the reverse shock could be as much as $\sim 10$ times the temperature of the shocked ejecta, but only at very early evolutionary stages ($\sim 1/5$ of the age in which the remnant becomes adiabatic). We note that if the cooler component is really to be associated with the ejecta, we would find that the metal abundances in the SN ejecta are below the cosmic values, because $Z_l = 0.24$. Even though the exact value of the abundances in the ejecta are difficult to predict, such a low value seems improbable. Vink et al. (1996) and Favata et al. (1997b), for instance, found that the low-T component in the X-ray spectrum of Cas A, which they have associated to the ejecta, has metal abundances well above the solar value.

If the SN ejecta expand in a medium in which substantial mass is present from circumstellar layers ejected before the SN explosion, the situation could be different. Interaction between ejecta and circumstellar medium (CSM) was discussed by Chevalier & Liang (1989) and shown to be consisted with observations of Cas A by Borkowski et al. (1996). The latters noted that in this case the high-T component may correspond to the SN ejecta, while the low-T component may correspond to the CSM. This scenario better fits to our data, because we find that the metallicity of the high-T component is above the cosmic value. The CSM abundances are expected to be approximately of solar-type, and Borkowski et al. (1996) found 0.40 for Fe and $\sim 0.9$ for the other metals.

If it is true that the X-ray emission of the SW shell of RCW86 is not associated to the ISM but to ejecta and CSM, the amount of shocked mass unrelated to the ISM is $< M_h + M_l = 4.2$ M$_\odot$ ($< 83$ M$_\odot$ if $E_{51} = 10$). This suggests that the progenitor SN is of Type Ia if $E_{51} = 1$ and of Type II if $E_{51} = 10$, in agreement with the conclusions reached in the previous section on the basis of the emission from the N shell.

Further insights on the presence of shocked ejecta can be provided by deep optical spectrophotometry. Leibowitz & Danziger (1983) examined the spectrum at different location of a very bright complex of filaments located around $14^h\ 40^m\ 20^s$ and $-62^d\ 39^m$, concluding that the abundances of iron should be lower than in Kepler, RCW103 and IC443. Their slit position falls in a region where the X-ray emission is soft and dominated by the $T_l$ component, therefore their results are in rough agreement with our measured $Z_l$. To properly address the topic of shocked ejecta, we need optical spectroscopy in regions dominated by the high-T component, where the optical surface brightness is much lower than the filament observed by Leibowitz & Danziger (1983).

Finding emission from the ejecta in RCW86 is crucial to the proper modeling of this object. In fact, if the connection of the remnant to the OB association at 2.8 kpc proposed by Rosado et al. (1996) is valid (at variance with the connection with SN185), then the solution with $E_{51} = 10$, $t = 4080$ yr and $M_{sw} = 157\ M_\odot$, presented in Table 7, must be favored. But, apart of the problem of a $E_{51}$ value much greater than the canonical value, this solution implies a very large swept-up ISM mass, and therefore it is not in agreement with the detection of ejecta in the SNR shell[2]. For this reason, on the basis of the X-ray data, we favor the solution which implies a relatively young SNR, $E_{51} = 1$, a Type Ia progenitor and the association with the historical supernova.

### 4.4. Spatial analysis of the two-temperature components in the SW region.

The hardness ratio maps presented in Fig. 4 provide a qualitative measurement of the spectral differences of the X-ray emission. Since the statistics and the BeppoSAX PSF do not allow us to investigate the observed regions with a full fit approach at angular scales smaller than several arcmin, we have tried to assess spatial spectral variations, which yield the hardness ratios in Fig. 4, with an alternative approach. This approach assumes that the observed spectral variations are entirely caused by variations of the ratio of emission measure $EM_l/EM_h$ of the two components detected in the SW and neglects that they could be due to temperature variations, or absorption effects. This is indeed the case of the Northern rim of the Vela shell, as pointed out by Bocchino et al. (1999), who have also developed a formalism which allows the filling factor of the cooler (or the hotter) plasma to be derived from the ratio $EM_l/EM_h$ and the density from the ratio $T_l/T_h$. Also Ozaki & Koyama (1997) performed spatially-resolved spectral analysis of the IC443 SNR and have found that all the SNR subregions investigated have

---

[2] We note however, that Miyata et al. (1998) detected metal-rich plasma in the central regions of the old Cygnus Loop SNR.

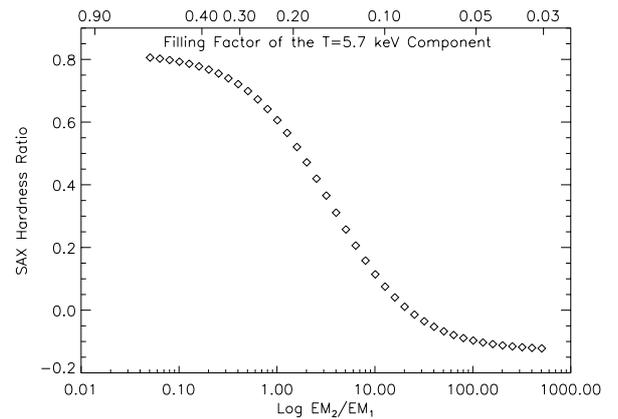

**Fig. 8.** Calibration curve of the SAX Hardness Ratio (defined as $(C_{2.0-10.5}^{\mathrm{MECS}} - C_{0.1-2.0}^{\mathrm{LECS}})/(C_{2.0-10.5}^{\mathrm{MECS}} + C_{0.1-2.0}^{\mathrm{LECS}})$, where $C$ are counts) versus the emission measure ratio $EM_l/EM_h$ assuming a two temperature thermal model with parameters fixed to values listed in Table 4

similar temperatures, the variations being mostly due to ionization time variations.

In order to model the hardness ratio in terms of the $EM_l/EM_h$ ratio, we have generated 40 simulated spectra with a two temperature model and parameter values fixed to the best-fit results listed in Table 4, column Var.Ab., except the value of the $EM_l/EM_h$ ratio, which spanned a range between 0.1 and 400 (the best-fit value is $\sim 3.5$). We have convolved the spectrum with the instrument response matrices and the effective area files which we have used for the spectral analysis of the LECS and MECS data. Then, for each spectrum we have computed the value of the ratio $(C_{2.0-10.5}^{\mathrm{MECS}} - C_{0.1-2.0}^{\mathrm{LECS}})/(C_{2.0-10.5}^{\mathrm{MECS}} + C_{0.1-2.0}^{\mathrm{LECS}})$, and we plotted it in Fig. 8 versus the input values of the emission measures ratio. The curve allows us to estimate the $EM_l/EM_h$ ratio from the hardness ratio maps. In Fig. 8, the top x-axis reports the value of the filling factors corresponding to the emission measure ratios of the bottom x-axis, computed using Eq. 4 of Bocchino et al. (1999) and assuming $kT_h = 5.7$ keV and $kT_l = 1.0$ keV.

Fig. 8 shows that hardness ratios (HRs) much greater than 0.85 and much lower than $-0.15$ cannot be produced in the framework of only varying the $EM_l/EM_h$ ratio. Fig. 9 is the histogram of the HR values reported in Fig. 4 and shows that only $\sim 15\%$ of the pixels has values outside this range. This is an indication that a two thermal component model with varying $EM_l/EM_h$ is a rather good description for most of the pixels. In the remaining $\sim 15\%$ of the pixels, there are probably additional effects, e.g. temperature variations. In fact, we note that the pixels having $HR > 0.85$ are mostly located at the inner edge of the dark region in Fig. 4, which is nearer than the "knee" to the center of the remnant, and therefore we expect an increase of the temperature there. In the "SW hard" region identified by Vink et al. (1997), the HR is in the range 0.5—0.9, corresponding to a filling factor $f_h > 0.20$, while in "SW soft" region we can identify two blobs where HR is between $-0.2$ and $0$ (labeled SW1 and SW2 in Fig. 4) and $f_h < 0.1$ and an "intermediate region" between them which connects the "SW Hard" region with the outside of the shell, in which HR is between $0$ and $0.5$ ($f_h = 0.1 - 0.2$). Therefore, most of the X-ray bright emission region in the "knee" is characterized by a low filling factor of the high temperature component, which becomes dominant only in the inner edge of this feature and probably in the central region of the remnant.

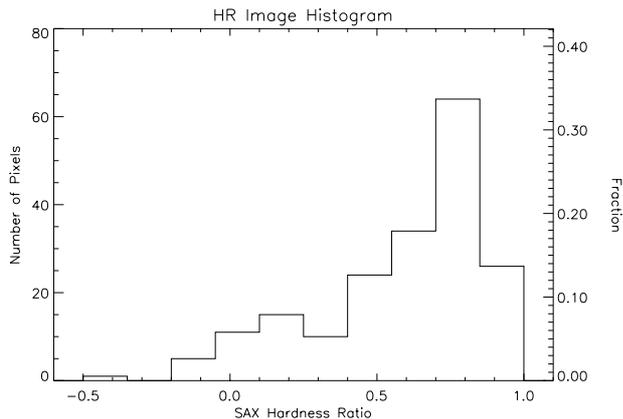

**Fig. 9.** Histogram of HR values reported in the Fig. 4. The size of the bin (0.15) is the typical uncertainty of HR

We note that the presence of the soft X-ray component at the edge of the rim favors the possibility that this component could be associated with the CSM, rather than to the ejecta.

## 5. Conclusions

The BeppoSAX observations of RCW 86 confirm that this remnant has different properties at different location in its shell, shedding more light on the nature of the differences. In particular, the N region shows X-ray properties indicating a Sedov shock interacting with the environment. The ROSAT spectrum is consistent with this picture and in addition seems to show a supersoft component ($kT = 0.2$ keV). If this component is real, we suggest that the Sedov-like properties correspond to the expansion inside a cavity wall, and that the shock has very recently encountered the wall of this cavity, giving rise to the supersoft component. Detailed spatially-resolved spectral analysis is required to exactly locate the supersoft component and to confirm this scenario. We have derived the abundances in this region and we have found that they are fully consistent with a metal-depleted ISM behind the shock in which grain destruction is occurring. Applying the Sedov analysis to the X-ray data, we find a distance of $1.18^{+0.17}_{-0.16}$ kpc and an age of $1630^{+440}_{-360}$ yr, if the initial explosion energy was $\sim 10^{51}$ erg. While we can probably rule out lower $E$, we cannot formally exclude a more energetic event on the basis of the North rim observations alone, which would require the remnant to be further away ($\sim 3$ kpc) and older ($\sim 4000$ yr).

The emission from the SW rim (the "knee") is very different, since two temperature components ($kT_h = 5.7$ and $kT_l = 1.0$ keV) are required to describe the data. Moreover the global metallicity of the high-T component ($Z_h$) are well above the cosmic abundances, suggesting that the hard SW emission mainly comes from stellar ejecta. The overall picture is consistent with the ejecta/CSM scenario discussed by Borkowski et al. (1996) for Cas A. The fact that different locations of the same SNR are described by very different models (as if they were different SNRs) is not really surprising, because it is well know that RCW 86 is a highly inhomogeneous object. We argue that spatial variations of the X-ray spectral emission of the "knee", as shown by the hardness ratio map in Fig. 4 (right panel), are well modeled by variations of the emission measure ratio of the two X-ray components, and, by using a formalism developed by Bocchino et al. (1999), we have estimated a filling factor of the high-T component $< 0.1$ in the soft X-ray regions and $0.2 - 0.9$ in the other SW regions. We have estimated an upper limit of the swept-up mass not directly related to the ISM, and we point out that the progenitor SN must be of Type Ia if we adopt the canonical $E_{51} = 1$. Moreover, the detection of the emission from ejecta favors the $E_{51} = 1$ solution in Table 7, which is compatible with the connection with SN185, thus weakening the also proposed connection of a more evolved RCW 86 with a more distant OB association at 2.8 kpc.

*Acknowledgements.* The BeppoSAX satellite is a joint Italian-Dutch programme. F. Bocchino acknowledges an ESA Fellowship, and thanks A. Parmar for useful criticisms and suggestions. F.B., A.M. and S.S. acknowledges partial support from Agenzia Spaziale Italiana and Ministero per l'Università e la Ricerca Scientifica.